\documentclass[letter]{aa}
\usepackage{txfonts}
\usepackage{graphicx}
\usepackage{xcolor}
\usepackage{multirow}
\usepackage{subcaption} 

\def\H2{H$_2$}

\begin{document}
\title{Deuterated methanol on Solar System scale around the HH212 protostar}

\author{E.Bianchi \inst{1,2} 
 \and 
C. Codella \inst{1}
\and
C. Ceccarelli \inst{3,4,1} 
\and
V. Taquet \inst{1}
\and 
S. Cabrit \inst{5}
\and
F. Bacciotti \inst{1}
\and
R. Bachiller \inst{6}
\and
E. Chapillon \inst{7,8}
\and
F. Gueth \inst{8}
\and 
A. Gusdorf \inst{9}
\and
B. Lefloch \inst{3,4}
\and
S. Leurini \inst{10}
\and 
L. Podio \inst{1}
\and 
K. L. J.  Rygl \inst{11} 
\and
B. Tabone \inst{5}
\and
M. Tafalla \inst{6}
 }

\institute{
INAF, Osservatorio Astrofisico di Arcetri, Largo E. Fermi 5,
50125 Firenze, Italy
\and
Universit\`a degli Studi di Firenze, Dipartimento di Fisica e Astronomia, 
Via G. Sansone 1, I-50019 Sesto Fiorentino, Italy
\and
Univ. Grenoble Alpes, Institut de
Plan\'etologie et d'Astrophysique de Grenoble (IPAG), 38401 Grenoble, France
\and
CNRS, Institut de
Plan\'etologie et d'Astrophysique de Grenoble (IPAG), 38401 Grenoble, France
\and
LERMA, Observatoire de Paris, PSL Research University, CNRS, Sorbonne Universit\'es, UPMC Univ. Paris 06,
\'Ecole Normale Sup\'erieure, 75014 Paris, France
\and
IGN, Observatorio Astron\'omico Nacional, Alfonso XII 3, 28014 Madrid, Spain
\and
Laboratoire d’astrophysique de Bordeaux, Univ. Bordeaux, CNRS, B18N, alle\'e Geoffroy Saint-Hilaire, 33615 Pessac, France
\and
IRAM, 300 rue de la Piscine, 38406 Saint-Martin-d’H\`eres, France
\and
LERMA, Observatoire de Paris, PSL Research University, CNRS, Sorbonne Universités, UPMC Univ. Paris 06, École normale supérieure, F-75005, Paris, France 
 \and
 INAF-Osservatorio Astronomico di Cagliari, Via della Scienza 5, I-09047, Selargius (CA), Italy
 \and
 INAF - Istituto di Radioastronomia \& Italian ALMA Regional Centre, Via P. Gobetti 101, I-40129 Bologna, Italy
}

\offprints{E. Bianchi, \email{ebianchi@arcetri.astro.it}}
\date{Received date; accepted date}

\authorrunning{Bianchi et al.}
\titlerunning{Deuterated methanol on Solar System scale around the HH212 protostar}
\abstract{Deuterium fractionation is a precious tool to understand the chemical evolution during the process leading to the formation of a Sun-like planetary system.}
{Methanol is thought to be mainly formed during the prestellar phase and its deuterated form keeps memory of the conditions at that epoch. Thanks to the unique combination of high angular resolution and sensitivity
provided by ALMA, we wish to measure methanol deuteration in the planet formation region around a Class 0 protostar and to understand its origin.}
{We mapped both the $^{13}$CH$_3$OH and CH$_2$DOH distribution in the inner regions ($\sim$100 au) of the HH212 system in Orion B. To this end, we used ALMA Cycle 1 and Cycle 4 observations in Band 7 with angular resolution down to $\sim$0$\farcs$15.} 
{We detected 6 lines of $^{13}$CH$_3$OH and 13 lines of CH$_2$DOH with upper level energies up to 438 K in temperature units.
We derived a rotational temperature of (171 $\pm$ 52) K and column densities of 7$\times$10$^{16}$ cm$^{-2}$ ($^{13}$CH$_3$OH) and 1$\times$10$^{17}$ cm$^{-2}$ (CH$_2$DOH), respectively. Consequently, the D/H ratio is (2.4 $\pm$ 0.4)$\times$10$^{-2}$, a value lower by an order of magnitude with respect to what was previously measured using
 single dish telescopes toward protostars located in Perseus.
Our findings are consistent with the higher dust temperatures
in Orion B with respect to that derived for the Perseus cloud.
The emission is tracing a
rotating structure extending up to 45 au from the jet axis
and elongated by 90 au along the jet axis. So far, the origin of the observed emission appears to be related with the accretion disk. Only higher spatial resolution measurements however, will be able to disentangle between different possible scenarios: disk wind, disk atmosphere, or accretion shocks.}
{}



\maketitle

\section{Introduction}

Molecular deuteration is a powerful diagnostic tool to study
the past history of the gas associated with the formation
of a proto-Sun and its protoplanetary system (see
e.g. Ceccarelli et al. 2014, and references therein).
In the prestellar core phase, the low temperatures 
and the resulting CO freeze-out enhance the
abundance of the deuterated molecules. These molecules are
stored in the icy grain mantles and then released into
the gas phase in the protostellar stage (Ceccarelli et al.
2007; Caselli et al. 2008).
This is the case of methanol (CH$_3$OH), formed on the
grain surfaces (e.g. Tielens 1983; Rimola et al. 2014)
 and then released into the gas phase because 
dust mantles are thermally evaporated in the
inner warm regions around low-mass protostars (e.g. Ceccarelli
et al. 2000, 2007; Parise et al. 2002, 2004, 2006) and/or sputtered by shocks
 (Codella et al. 2012; Fontani et al. 2014).

Deuterated methanol has been analysed towards
Class 0 protostars (Parise et al. 2006): the measured D/H of 0.4-0.6
indicates that CH$_{2}$DOH can be almost as
abundant as CH$_{3}$OH. 
In the only one more evolved source studied so far, the same ratio appears to be smaller
(D/H $\sim$ 1-7$\times$10$^{-3}$), 
suggesting a decrease of the deuterated species due to evolutionary effects
(Bianchi et al. 2017).
However, all the existing observations have been performed with single-dish
telescopes (angular resolution $\geq$ 10$\arcsec$)
which do not disentangle the
different components associated with a protostellar system
(jet, high-velocity shocks, slower accretion shocks, inner envelope).
Sub-arcsecond resolution observations are then crucial 
to measure the D/H ratio within 50–100 au from the protostar,
 i.e. the region hosting the protoplanetary disks.
Dust grains and ices may also be affected by the accretion shock, 
near the centrifugal barrier, so we do not know what the deuterium fractionation is at small scales.

The HH212 star-forming region in Orion ($d$ = 450 pc)
is an ideal laboratory to study the jet-disk system. 
The HH212-mm Class 0 protostar is driving a bright, extended,
and bipolar molecular jet extensively observed using IRAM-NOEMA, SMA, and
ALMA
(e.g. Lee et al. 2006, 2007, 2008, 2015, 2016, 2017a, 2017b; Codella et al. 2007;
Cabrit et al. 2007, 2012).
A molecular disk was revealed in
HCO$^+$, C$^{17}$O, and SO emission, using the first 
ALMA cycles with angular resolutions of $\sim$ 0$\farcs$6.
Velocity gradients have been detected
along the equatorial plane consistently with a rotating
disk around a 0.2-0.3 M$_{\rm \odot}$
protostar (Lee et al. 2014; Codella et al. 2014; Podio et al. 2015).
Indeed, a disk with a radius of 60 au has been imaged 
by Lee et al. (2017a)
observing dust continuum emission at 850 $\mu$m. 
Finally, Codella et al. (2016) and Leurini et al. (2016)
suggested that HDO and CH$_3$OH emission is likely associated with 
outflowing gas, and possibly with a disk wind.
To conclude, the HH212 region is the unique 
Class 0 protostellar region where
a bright jet, a compact rotating disk, and signatures of 
a disk wind have been revealed.
In this Letter, we
exploit ALMA Band 7 observations with an angular resolution
down to 0$\farcs$15 to obtain the first measurement of 
methanol deuteration in the disk formation region.

\section{Observations} 

HH212 was observed in Band 7
with ALMA using 34 12-m antennas  
between 15 June and 19 July 2014 during the Cycle 1 phase.
Further observations of HH212 were performed in
Band 7 with ALMA using 44 12-m antennas  
between 6 October and 26 November 2016 during the Cycle 4 phase.
The maximum baselines for Cycle 1 and 4 were 650 m and 3 km, respectively.

In Cycle 1 the spectral windows between 337.1--338.9 GHz and 348.4--350.7 GHz 
were observed using spectral channels of 488 kHz (0.42--0.43 km s$^{-1}$), 
subsequently smoothed to 1.0 km s$^{-1}$.
Calibration was carried out following standard procedures, 
using quasars J0607-0834, J0541--0541, J0423--013, and Ganymede.
The continuum-subtracted images have
clean-beam FWHMs from $0\farcs41\times0\farcs33$ to
$0\farcs52\times0\farcs34$
(PA = -63$\degr$), and an rms noise level of
5--6 mJy beam$^{-1}$ in 1.0 km s$^{-1}$ channels.  
 For Cycle 4 spectral units of 977 kHz (0.87 km s$^{-1}$) 
were used to observe the spectral windows between 334.1--336.0 GHz.
Calibration was carried out following standard procedures, 
using quasars J0510+1800, J0552+0313, J0541--0211 and J0552--3627.
 The continuum-subtracted images have a typical 
clean-beam FWHM of $0\farcs15\times0\farcs12$
(PA = -88$\degr$), and an rms noise level of
1 mJy beam$^{-1}$ in 0.87 km s$^{-1}$ channels.
Spectral line imaging was achieved with the CASA package, while 
data analysis was performed using the 
GILDAS\footnote{http://www.iram.fr/IRAMFR/GILDAS} package. 
Positions are given with respect to the MM1 protostar continuum peak
located at $\alpha({\rm J2000})$ = 05$^h$ 43$^m$ 51$\fs$41, 
$\delta({\rm J2000})$ = --01$\degr$ 02$\arcmin$ 53$\farcs$17
(Lee et al. 2014).
Lines were identified using 
spectroscopic parameters extracted from the Jet Propulsor Laboratory  
(JPL\footnote{https://spec.jpl.nasa.gov/}, Pickett et al. 1998) and Cologne Database
 for Molecular Spectroscopy (CDMS\footnote{http://www.astro.uni-koeln.de/cdms/}; 
 M\"uller et al. 2001, 2005) molecular databases.
  
\begin{figure}[h]
\centerline{\includegraphics[angle=0,width=9cm]{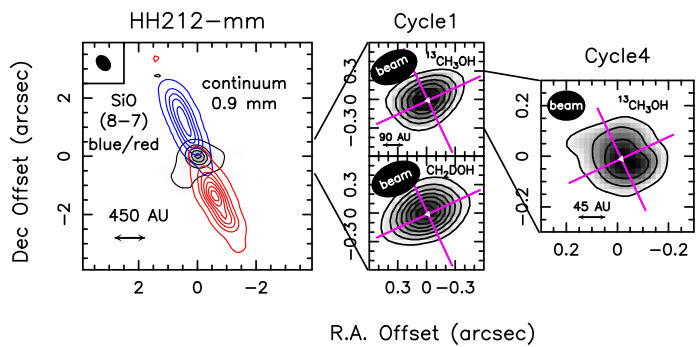}}
\captionsetup{labelfont=bf}
\caption{{\it Left Panel:} The HH212 protostellar system
as observed by ALMA-Band 7 (Codella et al. 2014).
Blue/red contours plot the blue/red-shifted SiO(8--7) jet
overlaid on the continuum at 0.9 mm (black contours).
Positions are with respect to the coordinates
reported in Sect. 2.
The filled ellipse shows
the synthesised beam (HPBW):
$0\farcs61\times0\farcs45$.
{\it Central Panels:} Zoom-in of the central region
as observed by ALMA-Band 7 Cycle 1:
$^{13}$CH$_3$OH(13$_{0,13}$--12$_{1,12}$)E and
CH$_{2}$DOH(9$_{0,9}$--8$_{1,8}$)e0 integrated
over $\pm$4 km s$^{-1}$ with respect to the $v_{sys}$ = +1.7
km s$^{-1}$, (black contours and grey scale).
First contours and steps are 5$\sigma$
for $^{13}$CH$_3$OH (15 mJy beam $^{-1}$ km s$^{-1}$) and
CH$_{2}$DOH (35 mJy beam $^{-1}$ km s$^{-1}$).
The HPBWs are: $0\farcs52\times0\farcs34$ (PA = -64$\degr$) for
$^{13}$CH$_3$OH and $0\farcs54\times0\farcs35$ (PA = -65$\degr$) for
CH$_{2}$DOH.
{\it Right Panel:}
Zoom-in of the central region, as observed by ALMA-Band 7 Cycle 4,
showing the $^{13}$CH$_3$OH(12$_{1,11}$--12$_{0,12}$)A
emission integrated over 10 km s$^{-1}$ around $v_{sys}$
(black contours and grey scale).
First contours and steps are 5$\sigma$
(20 mJy beam $^{-1}$ km s$^{-1}$) and 3$\sigma$, respectively.
The HPBW is $0\farcs15\times0\farcs12$ (PA = -88$\degr$).}
\label{maps}
\end{figure}
 
\section{Results and Discussion}

\subsection{Line spectra and maps}

In Cycle 1 data, we detected 5 lines of $^{13}$CH$_{3}$OH and 
12 lines of CH$_{2}$DOH covering
excitation energies, $E_{\rm up}$, from 17 to 438 K (see Table \ref{tab}).
In addition, in the Cycle 4 dataset we revealed one transition of $^{13}$CH$_{3}$OH ($E_{\rm up}$= 193 K), and
one of CH$_{2}$DOH ($E_{\rm up}$= 327 K).
Examples of the profiles of the detected lines are shown in Fig. \ref{spectra}: the
emission peaks at velocities between $\simeq$ +1 km s$^{-1}$ and
+3 km s$^{-1}$, consistently with the systemic velocity of 
+1.7 km s$^{-1}$ (Lee et al. 2014), once considered the 
 spectral resolution of 1 km s$^{-1}$ .
The profiles, fit using the GILDAS package, are Gaussian with typical Full 
Width at Half Maximum
 (FWHM) $\sim$ 5 km s$^{-1}$.
The spectral parameters of the detected lines are presented in Table \ref{tab}.
Figure \ref{maps} introduces the HH212 system: the 0.9 mm continuum 
 traces the envelope hosting the protostar. The bipolar blue- 
 and red-shifted lobes are traced by SiO (Codella et al. 2007; Lee et al. 2014).
The central panels, covering a zoom-in of the central protostellar region 
($\lesssim 0\farcs5$, i.e. 225 au), show the Cycle 1 data, 
specifically examples of the emission maps 
of $^{13}$CH$_{3}$OH and CH$_{2}$DOH.
The emission is spatially unresolved with a size smaller than the Cycle 1 beam ($0\farcs5 \times 0\farcs3$)
corresponding to a size of 225$\times$135 au. 
However, the definitely higher angular resolution provided by
the Cycle 4 dataset ($0\farcs15 \times 0\farcs12$ corresponding to a radius
of 34$\times$27 au) provides spatially resolved images
of both $^{13}$CH$_{3}$OH and CH$_{2}$DOH\footnote{The 
CH$_{2}$DOH line is blended with the
SO$_2$(8$_{2,6}$--7$_{1,7}$) line (Tab. \ref{tab}).} emission lines.  
We used the $^{13}$CH$_{3}$OH emission as revealed in the Cycle 4 image to
show in the right panel of Fig. \ref{maps} a further zoom-in,
sampling the inner 100 au. 

 \begin{figure*}
\centerline{\includegraphics[angle=0,width=16cm]{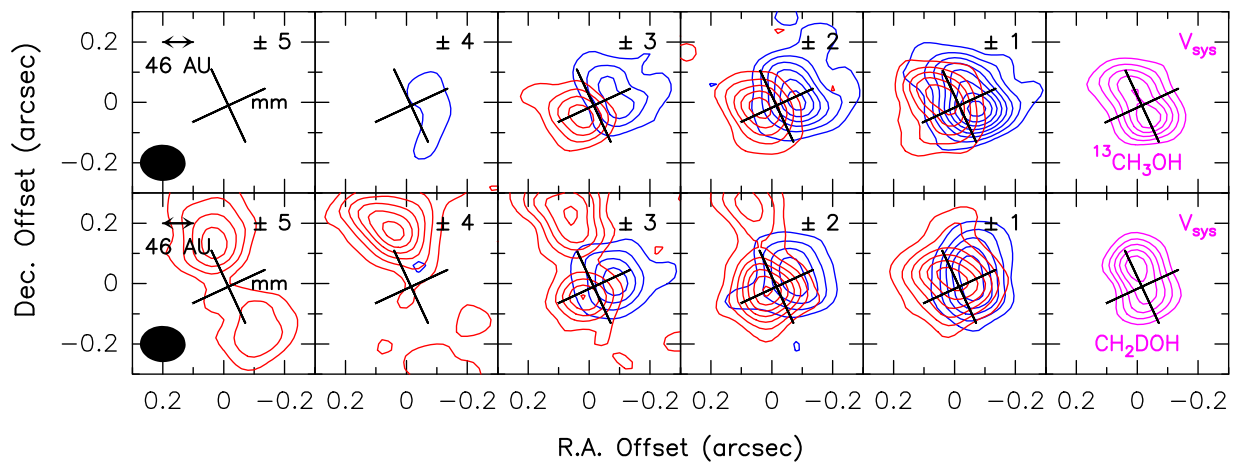}}
\captionsetup{labelfont=bf}
\caption{Channel maps of the $^{13}$CH$_3$OH(12$_{1,11}$--12$_{0,12}$)A
(upper panels) and CH$_2$DOH(16$_{2,15}$--16$_{1,15}$)o1 (lower panels)
blue- and redshifted
(continuum subtracted) emissions observed during ALMA-Cycle 4
towards the HH212-mm protostar.
Each panel shows the emission integrated over a velocity interval
of 1 km s$^{-1}$ shifted with respect to
the systemic velocity (see the magenta channel, sampling the velocity
between +1.5 km s$^{-1}$ and +2.5 km s$^{-1}$) by the
value given in the upper-right corner.
The black cross (inclined in order to point the SiO jet direction and
consequently the equatorial plane, see Fig. 1)
indicates the position of the protostar.
The ellipse in the top left panel shows
the ALMA synthesised beam (HPBW): $0\farcs15\times0\farcs12$ (PA = --88$\degr$).
First contours and steps correspond to 3$\sigma$ (2.7 mJy beam $^{-1}$ km s$^{-1}$). 
Note that for CH$_2$DOH at velocities larger than +1.0 km s$^{-1}$
a further emission appears along the jet axes to the north and to the south with respect to the 
CH$_2$DOH emission. This is due to the
SO$_2$(8$_{2,6}$--7$_{1,7}$) emission line (see Tab. 1 and Fig. 2),
which confirms to be a good tracer of outflowing matherial (Podio et al. 2015).
Given that the velocity scale has been derived by
first fixing the CH$_{2}$DOH frequency, the SO$_{2}$ emission looks artificially red, but it is, obviously,
blue-shifted (see Fig. A.1).}
\label{channels}
\end{figure*}

\subsection{Methanol deuteration}

We used the $^{13}$CH$_{3}$OH lines detected in Cycle 1 to perform 
a rotation diagram analysis, assuming LTE conditions
and optically thin lines. 
In order to exclude opacity effects,
we use the $^{13}$CH$_{3}$OH lines to derive the CH$_{3}$OH
column density as the main isotopologue lines observed by Leurini et al. (2016) 
are moderately optically thick ($\tau <$ 0.4). 
We also verified that the CH$_{2}$DOH transitions are indeed optically thin
using a Population Diagram analysis (Goldsmith \& Langer 1999)
in which opacities are self-consistently computed. 
We used $^{13}$CH$_{3}$OH  and CH$_{2}$DOH lines extracted from 
the same ALMA dataset in order to have the same $u-v$ coverage and to minimise 
calibration effects.
Moreover, the Cycle 0 (beam $\simeq$ 0$\farcs$6) observations of CH$_3$OH 
(Leurini et al. 2016) trace a much warmer component 
dominated by lines with $E_{\rm up}$ up to 747 K with respect to the emission lines analysed in this paper.
A source size of 0$\farcs$19 $\pm$ 0$\farcs$02 has been derived
from the $^{13}$CH$_{3}$OH Cycle 4 map integrated over
the whole emission range and then used to correct the Cycle 1 observed values.
This value is in good agreement with $\sim$ 0$\farcs$2, as
derived from CH$_{3}$OH (Leurini et al. 2016).
Figure \ref{RD} shows the derived rotational temperature and column density which are
$T_{\rm rot}$ = (171 $\pm$ 52) K and $N_{\rm tot}$ = (6.5 $\pm$ 2.1) $\times$ 10$^{16}$ cm$^{-2}$,
respectively. 
Conservatively, we used only Cycle 1 data to perform the fit; 
however, the $^{13}$CH$_{3}$OH Cycle 4 line is in 
good agreement with Cycle 1 observations.
In the lower panel of Fig. \ref{RD} we derive the column density of 
CH$_{2}$DOH, assuming the same rotational temperature and size derived for $^{13}$CH$_{3}$OH.
We excluded from the fitting the two 
CH$_{2}$DOH transitions with $E_{\rm up}$ $>$ 300 K to exclude
any possible contamination from non-thermal excitation processes or the occurrence of a component
with a different, higher, excitation condition.
However the fitting obtained with the low-energies transitions (continuous line) is in agreement
with the two excluded transitions, as well as with the Cycle 4 transition,
as shown by the dotted line. The derived column density for 
CH$_{2}$DOH is then $N_{\rm tot}$ = (1.1 $\pm$ 0.2) $\times$ 10$^{17}$ cm$^{-2}$. 
Note that an LVG analysis (see Ceccarelli et al. 2003 for further details) 
of the $^{13}$CH$_{3}$OH emission indicates a
source size of $0\farcs1$-$0\farcs2$, $T_{\rm kin}$ larger than 100 K and densities larger than 10$^{6}$ cm$^{-3}$ in agreement with the HDO results reported by Codella et al. (2016).

The column densities derived from the rotation diagrams have been used to
derive the methanol D/H for the first time in the inner 100 au
around a low-mass protostar. Assuming a $^{12}$C/$^{13}$C ratio of 70
(Milam et al. 2005)  at the galactocentric distance of HH212 ($\sim$ 8.3 kpc), the D/H is 
consequently (2.4 $\pm$ 0.4) $\times$ 10$^{-2}$.  
This value is in agreement with the upper limit of 0.27 derived by Lee et al. (2017b)
using CH$_{2}$DOH and three lines of CH$_{3}$OH. In addition, our measurement
 is lower than previous onces of CH$_{2}$DOH 
in other Class 0 objects performed with the IRAM-30m
single dish, which indicate D/H $\simeq$ 40-60$\%$  (Parise et al. 2006).
Taking into account that, as reported by Belloche et al. (2016), the column densities
of CH$_{2}$DOH derived in Parise et al. (2006) were overestimated by
a factor $\sim$2 because of a problem in the spectroscopic parameters,
the HH212 deuteration is still one order of magnitude lower. 
Which are the reasons of this difference$?$
To start with, the single-dish measurements by Parise et al. (2006) are
sampling regions $\geq$ 2000 au: the D/H is calculated assuming that the main isotopologue
and the deuterated species come from the same emitting source, 
but this cannot be verified with low angular resolutions.
This issue is overcome with high angular resolution observations, like those presented
here that allow us to directly image the emitting region.
Interestingly, J\o{}rgensen et al. (2016) report a
level of deuteration for glycolaldehyde of $\sim$5 $\%$ in IRAS16293 
in Ophiuchus on 50 au scale.
That said, a possible explanation for the lower methanol deuteration 
measured for HH212-mm could be related to different physical conditions,
 during the formation of methanol on
dust mantles during the prestellar phase.
Indeed, a larger gas temperature reduces the atomic gas D/H ratio landing on the grain surfaces, 
reducing, consequently, the deuteration of methanol.
All the sources observed by Parise et al. (2006) are located 
in the Perseus star-forming region,
which could had experienced different conditions with respect to
the Orion B region where HH212 lies.
Specifically, while in Perseus the dust temperature is about $\sim$12 K (Zari et al. 2016),
HH212 is located about one degree north of the high-mass star forming
 region NGC 2024, and the dust temperature here is $\geq$16 K (Lombardi et al. 2014).
As shown e.g. by Taquet et al. (2012, 2014), deuteration (including that of methanol),
once the volume density is fixed, decreases as temperature increases. The models
 indicate that D/H can decrease by up to
one order of magnitude by increasing the temperature from 10 K to 20 K.
Interestingly, we note that
according to Fuente et al. (2014) the methanol deuteration
in the hot core in the intermediate mass star forming  region NGC7129, FIRS 2
is $\sim$ 2 $\%$.
However in hot cores associated with massive 
star forming regions in Orion (Peng et al. 2012),
the methanol D/H is between 0.8 $\times$ 10$^{-3}$ and 1.3 $\times$ 10$^{-3}$, lower by one order of
magnitude than the values reported here.

\subsection{The emitting region of deuterated methanol}

Figure \ref{channels} shows the channel maps of the two transitions observed during ALMA Cycle 4, i.e. at the highest
spatial resolution presented here: $^{13}$CH$_{3}$OH (12$_{\rm 1,11}$--12$_{\rm 0,12}$)A and 
CH$_{2}$DOH (16$_{\rm 2,15}$--16$_{\rm 1,15}$)o1 (see Table \ref{tab}). For both lines
the rightmost panels show the spatial distributions
of the emission close to systemic velocity.
The rest of the panels are for the blue- and red-shifted velocities, imaged up to
$\pm$4 km s$^{-1}$. Note that for CH$_{2}$DOH 
the red-shifted emission is partially contaminated by the
SO$_2$(8$_{\rm 2,6}$--7$_{\rm 1,7}$) emission (see Table \ref{tab} and Fig. \ref{spectra}).

The spatial distribution of the methanol isotopologues at the systemic velocity
is elongated along the jet axis (PA = 22$\degr$).
A fit in the image plane gives for $^{13}$CH$_{3}$OH a beam deconvolved FWHM size
of 0$\farcs$18(0$\farcs$02)$\times$0$\farcs$12(0$\farcs$02) at PA = 33$\degr$.
For CH$_{2}$DOH we derived 0$\farcs$20(0$\farcs$02)$\times$0$\farcs$10(0$\farcs$02) 
and PA = 19$\degr$.
These values correspond to FWHM sizes of 81$\times$54 au ($^{13}$CH$_{3}$OH)
and 90$\times$45 au (CH$_{2}$DOH). 
Moving towards the blue- and red-shifted emission we 
resolve a clear velocity gradient parallel to the equatorial plane (see also Fig. \ref{spectra-blue-red})
with shifts of $\pm$ 0$\farcs$1 = 45 au, in the same sense as the HH212 disk rotation. 
The  typical (beam deconvolved) size of the blue- and red-shifted emission 
at $\pm$2 km s$^{-1}$ from systemic
is 0$\farcs$22(0$\farcs$04)$\times$0$\farcs$14(0$\farcs$04)
and 0$\farcs$18(0$\farcs$02)$\times$0$\farcs$10(0$\farcs$03), respectively,
similar to the size at V$_{\rm sys}$.
Rotation was already noted in CH$_3$OH by Leurini et al. (2016) 
from unresolved maps at lower resolution ($\simeq$ 0$\farcs$6).
The authors further noted that emission centroids (fitted in $u-v$ plane) 
moved away from the source, and eventually away from the mid plane, at higher velocity, 
speculating that the methanol emission is not dominated by a
Keplerian disk or the rotating-infalling cavity, and possibly associated with a disk wind.
With the present images we reach higher angular resolution 
which allows us to refine the picture: the systemic velocity looks arising mainly from two peaks at $\pm$0$\farcs$05 = 20 au above and below the disk plane
(of which the centroids in Leurini et al. 2016 only traced the barycenter).
 In addition, channel maps at $\pm$1 km s$^{-1}$
suggest red-shifted emission peaking in the north and blue-shifted
emission peaking in the south,
which is not the sense we would naively expect for a wind from HH212-mm.
Since the jet axis is so close to the plane of the sky\footnote{The system inclination is 4$\degr$ to the plane of sky; Claussen et al. 1998)},
this might perhaps still be explained by non-axisymmetric structure in a disk wind,
but synthetic predictions would be needed to test this. 
Alternatively, our data suggest that the emission could arise from
accretion shocks occurring at the centrifugal barrier associated with
the accretion disk, and heating the gas at temperatures around 100 K
(Sakai et al. 2014; Oya et al. 2016). 
Indeed, the radial extent $\sim$ 50 au
of the observed red- and blue-shifted emission 
is consistent with the dust disk radius $\sim$ 60 au, recently resolved 
by Lee et al. (2017a) at 850 $\mu$m with ALMA.
In this case, we would observe a thicker accretion shock near the centrifugal barrier in HH212, as recently seen
in C$_2$H in L1527 (Sakai et al. 2017).
We may also have a contribution from a 
warm disk atmosphere, emitting in the northern/southern portions of the accretion disk,
thicker than the dust atmosphere.


\section{Conclusions}

The ALMA Cycle 1 and 4 observations allow us to measure
methanol
deuteration in the inner 50 au of the jet-disk system associated
with the Class 0 protostar HH212, in Orion B. The deuteration is
$\simeq$
2 $\times$ 10$^{-2}$, a value lower than what
previously measured using single-dish towards Class 0
protostars in Perseus. 
Although we cannot exclude that single-dish
observations are mixing different gas components with different D/H
values,
our findings are consistent with a higher dust temperature
in Orion B with respect to the Perseus cloud.
This confirms the diagnostic value of molecular deuteration to recover the physical conditions during the pre-collapse phase.
The emission is confined in a
rotating structure which extends at $\pm$ 45 au from the equatorial
plane and is elongated along the jet axis. Disk wind, disk atmosphere, and accretion shocks could explain the 
observed images. Higher spatial resolution maps will be necessary to distinguish between these possibilities.

\clearpage

\appendix
\begin{acknowledgements}
We thank C.-F. Lee for instructive comments and suggestions.
We also thank the anonymous referee for having improved the manuscript.
This paper makes use of the ADS/JAO.ALMA\#2012.1.00997.S and ADS/JAO.ALMA\#2016.1.01475.S data (PI: C. Codella). ALMA is a partnership of 
ESO (representing its member states), NSF (USA) and NINS (Japan), together with NRC (Canada) and NSC 
and ASIAA (Taiwan), in cooperation with the Republic of Chile. The Joint ALMA Observatory is operated by ESO, AUI/NRAO and NAOJ.
This work was supported 
by (i) by the Italian Ministero dell'Istruzione, Universit\`a e Ricerca (MIUR) through 
the grant Progetti Premiali 2012 - iALMA (CUP C52I13000140001), and
(ii) by the program PRIN-MIUR 2015 STARS in the CAOS (Simulation Tools 
for Astrochemical Reactivity and Spectroscopy in the Cyberinfrastructure 
for Astrochemical Organic Species, 2015F59J3R, MIUR e della 
Scuola Normale Superiore).
M.T. aknowledges partial support from the project AYA2016-79006-P.
\end{acknowledgements}

\section{Additional material}

\begin{figure}
\centerline{\includegraphics[angle=0,width=9cm]{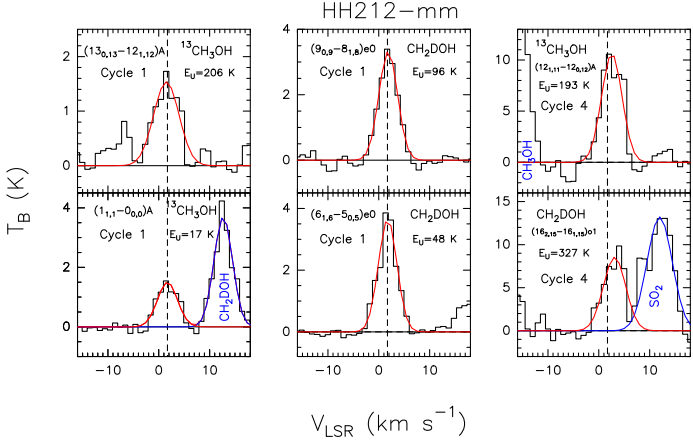}}
\captionsetup{labelfont=bf}
\caption{Examples of line profiles in T$_B$ scale (not corrected for the beam dilution): species and transitions are reported. The vertical dashed line stands for the systemic V$_{LSR}$ velocity (+ 1.7 km s$^{-1}$, Lee et al. 2014). Red and blue curves are for the Gaussian fit. In the spectrum of Cycle 1 (1$_{1,1}$--0$_{0,0}$)A $^{13}$CH$_{3}$OH transition, a CH$_{2}$DOH doublet containing  the (5$_{4,2}$--5$_{3,2}$)e1 and the (5$_{4,1}$--5$_{3,3}$)e1 transitions is also shown (see Table \ref{tab}).
In Cycle 4 spectra the (7$_{1,7}$--6$_{1,6}$)A CH$_{3}$OH 
and the (8$_{2,6}$--7$_{1,7}$) SO$_{2}$ transitions are also shown.}

\label{spectra}
\end{figure}

\begin{figure}[htp]
\begin{center}
    \begin{subfigure}{0.4\textwidth}

        \includegraphics[scale=0.34]{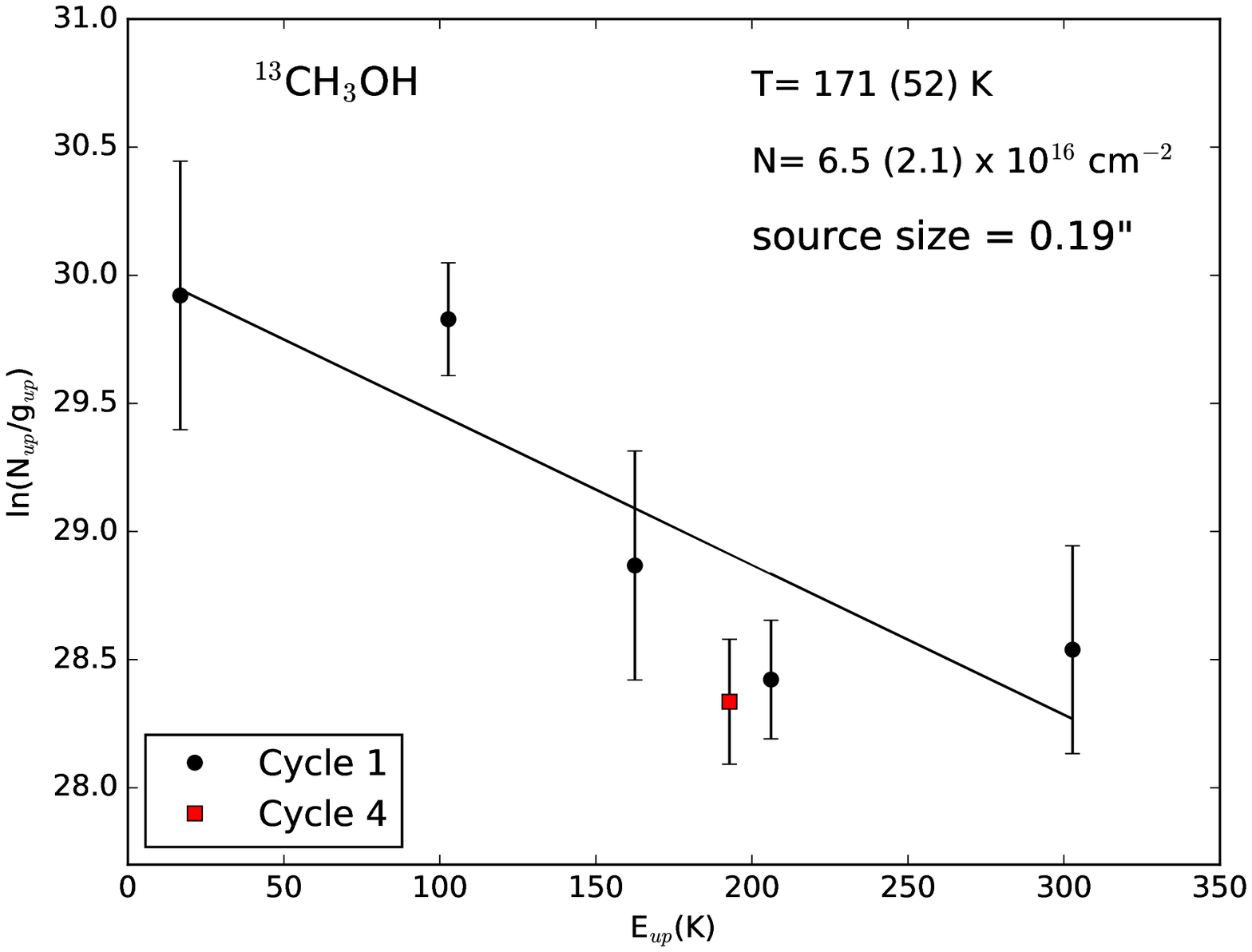}

    \end{subfigure}

    \hspace{2 cm}
    \begin{subfigure}{0.4\textwidth}
\includegraphics[scale=0.34]{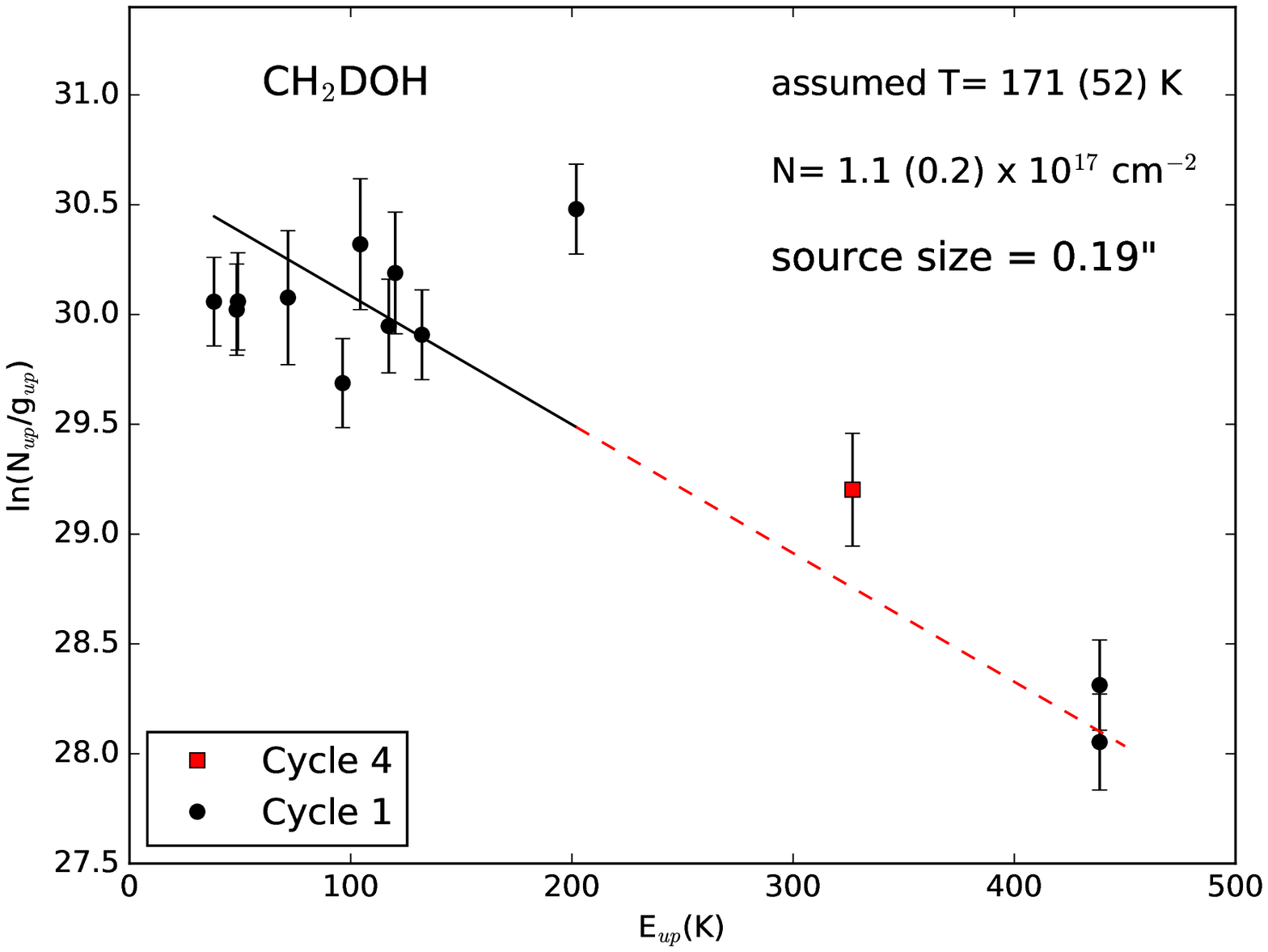}

  \end{subfigure}
  \end{center}
    \captionsetup{labelfont=bf}
    \caption{Rotation diagrams for $^{13}$CH$_{3}$OH (upper panel) and CH$_2$DOH (lower panel). An emitting region size of $0\farcs19$ is assumed (see text). The parameters $N_{\rm u}$, $g_{\rm u}$, and $E_{\rm up}$ are, respectively, the column density, the degeneracy and the energy of the upper level. The derived value of the rotational temperature is reported for $^{13}$CH$_{3}$OH. For CH$_2$DOH the $T_{\rm rot}$ derived from the $^{13}$C-isotopologues is assumed to derive the column density. The two CH$_2$DOH transitions with excitation energies higher than 400 K as well as the Cycle 4 point are excluded from the fit although they are in good agreement with it.} 
\label{RD}
\end{figure}

\begin{figure}[h]
\centerline{\includegraphics[angle=0,width=8cm]{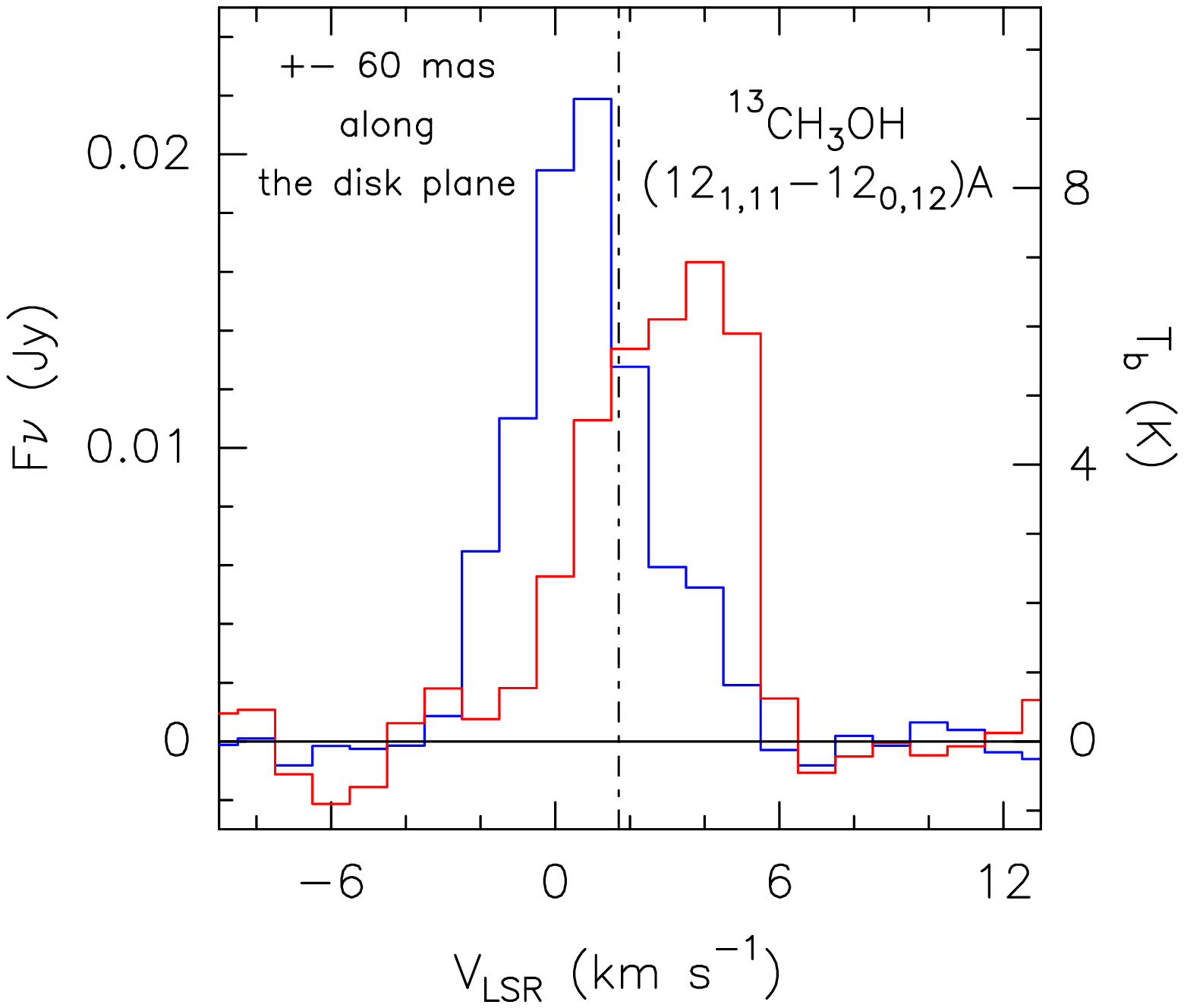}}
\captionsetup{labelfont=bf}
\caption{Comparison in flux density and in $T_{\rm B}$ scales ($T_{\rm B}$/$F_{\rm \nu}$ = 482.609 K Jy$^{-1}$)
between the $^{13}$CH$_{3}$OH(12$_{\rm 1,11}$--12$_{\rm 0,12}$)A
spectrum extracted at $\pm$ 0$\farcs$06 from the protostar (see Fig. 2:
blue- and red-shifted emission towards North-West and South-East, respectively)
in the direction
along the equatorial plane (i.e. disk plane; Lee et al. 2017). The vertical dashed line stands for the systemic V$_{LSR}$ velocity (+ 1.7 km s$^{-1}$, Lee et al. 2014).}
\label{spectra-blue-red}
\end{figure}

Table \ref{tab} lists all the methanol isotopologue lines observed towards HH212-mm during ALMA Cycle 1 and Cycle 4 operations.
Fig. \ref{spectra} shows examples of line profiles in T$_B$ scale, while
Fig. \ref{RD} shows the rotation diagrams for $^{13}$CH$_{3}$OH and CH$_2$DOH. 
Fig. \ref{spectra-blue-red} shows the $^{13}$CH$_{3}$OH(12$_{\rm 1,11}$--12$_{\rm 0,12}$)A
spectrum extracted at $\pm$ 0$\farcs$06 from the protostar, in the equatorial plane direction.

\begin{table} 
 \caption{Emission lines detected in ALMA Cycle 1 and 4 measurements towards HH212-mm.}
\label{tab}
\begin{small}

\begin{tabular}{lccccc}

\hline
\hline
\multicolumn{1}{c}{Transition} &
\multicolumn{1}{c}{$\nu$$^{\rm a}$} &
\multicolumn{1}{c}{$E_{\rm up}$$^a$} &
\multicolumn{1}{c}{$S\mu^2$$^a$} &
\multicolumn{1}{c}{rms} &
\multicolumn{1}{c}{$F_{\rm int}$$^b$} \\
\multicolumn{1}{c}{ } &
\multicolumn{1}{c}{(GHz)} &
\multicolumn{1}{c}{(K)} &
\multicolumn{1}{c}{(D$^2$)} & 
\multicolumn{1}{c}{(K)} &
\multicolumn{1}{c}{(K km s$^{-1}$)} \\ 

\hline
\multicolumn{6}{c}{$^{13}$CH$_{\rm 3}$OH  Cycle 1}\\
\hline


 13$_{\rm 0,13}$--12$_{\rm 1,12}$A & 338.75995 & 206 & 13 & 0.29 & 9.6(1.1)\\

 13$_{\rm 4,9}$--14$_{\rm 3,11}$E & 347.78840 & 303 & 4 & 0.36 &  4.0(1.4)\\

 11$_{\rm 0,11}$--10$_{\rm 1,9}$E & 348.10019 & 162 & 5 & 0.84 & 7.5(3.0)\\

 1$_{\rm 1,1}$--0$_{\rm 0,0}$A & 350.10312 & 17 & 2 &  1.16 &  7.4(3.0)\\

 8$_{\rm 1,7}$--7$_{\rm 2,5}$E & 350.42158 & 103 & 2 & 0.93 &  8.7(2.6)\\



\hline
\multicolumn{6}{c}{CH$_{\rm 2}$DOH   Cycle 1}\\
\hline

  9$_{\rm 0,9}$--8$_{\rm 1,8}$ e0 & 337.34866 & 96 & 6 & 0.15 &15.3(0.5) \\

  9$_{\rm 1,8}$--8$_{\rm 2,6}$ o1$^c$ & 338.46254 & 120 & 2 & 0.26 & 10.7(2.1)$^c$ \\

  13$_{\rm 1,12}$--12$_{\rm 0,12}$ e0 & 338.86898 & 202 & 2 &  0.13 & 9.7(0.4) \\

  6$_{\rm 1,6}$--5$_{\rm 0,5}$ e0 & 338.95711 & 48 & 5 &  0.29 & 16.7(0.9) \\

  18$_{\rm 4,15}$--18$_{\rm 3,15}$ e1 & 347.76728 & 438 & 10 &  0.09 &  6.6(0.3) \\

 18$_{\rm 4,14}$--18$_{\rm 3,16}$ e1 & 347.95281 & 438 & 10 &  0.18 &  6.4(0.6) \\
\vspace{0.1 cm}
 4$_{\rm 1,3}$--4$_{\rm 0,4}$ e1 & 348.16076 & 38 & 4 &    0.15 & 18.1(0.5) \\

  7$_{\rm 4,4}$--7$_{\rm 3,4}$ e1 & 349.95168 &  132 & 3 &  \multirow{2}{*}{0.16} & \multirow{2}{*}{25.3(0.5)} \\
\vspace{0.1 cm}
  7$_{\rm 4,3}$--7$_{\rm 3,5}$ e1 & 349.95272 &  132 & 3 & &   \\

  6$_{\rm 4,3}$--6$_{\rm 3,3}$ e1 & 350.02735 &  117 & 2 & \multirow{2}{*}{0.23} &\multirow{2}{*}{19.5(0.7)}  \\
\vspace{0.1 cm}
 6$_{\rm 4,2}$--6$_{\rm 3,4}$ e1 & 350.02777 &  117 & 2 &  &    \\

  5$_{\rm 4,2}$--5$_{\rm 3,2}$ e1 & 350.09024 &  104 & 2 & \multirow{2}{*}{0.63}  &\multirow{2}{*}{20.0(2.2)}  \\
\vspace{0.1 cm}
  5$_{\rm 4,1}$--5$_{\rm 3,3}$ e1 & 350.09038 &  104 & 2 & &  \\

  6$_{\rm 2,5}$--5$_{\rm 1,5}$ e1$^c$ & 350.45387 &  72 & 4 & 0.63 &  17.0(3.9)$^c$ \\

  5$_{\rm 1,4}$--5$_{\rm 0,5}$ e1 & 350.63207 &  49 & 5 & 0.60 &  21.0(2.0) \\

\hline
\multicolumn{6}{c}{$^{13}$CH$_{\rm 3}$OH Cycle 4}\\
\hline


 12$_{\rm 1,11}$--12$_{\rm 0,12}$ A & 335.56021 & 193 & 23 & 0.81 &  61.7(2.7) \\

\hline
\multicolumn{6}{c}{CH$_{\rm 2}$DOH Cycle 4}\\
\hline

  16$_{\rm 2,15}$--16$_{\rm 1,15}$ o1$^d$ & 334.68395 &  327 & 8 & 0.67  &  51.1(2.9)$^d$  \\

\hline

\end{tabular}

\end{small}
\vspace{0.5cm}
{\bf Notes}. $^a$ Frequencies and spectroscopic parameters are extracted from the Jet Propulsion Laboratory molecular database (JPL, Pickett et al. 1998) and from the Cologne Database for Molecular Spectroscopy (CDMS, Müller et al. 2001). Upper level energies refer to the ground state of each symmetry. $^b$ Gaussian fit. 
$^c$ The line transition (9$_{\rm 1,8}$--8$_{\rm 2,6}$)o1 is partially blended with the (7$_{\rm -5,2}$--6$_{\rm -5,1}$)E1 CH$_{\rm 3}$OH transition. The line (6$_{\rm 2,5}$--5$_{\rm 1,5}$)e1 is partially blended with the (18$_{\rm 1,17}$--17$_{\rm 1,16}$)E CH$_{\rm 3}$CHO transition. $^d$ The CH$_{\rm 2}$DOH (16$_{\rm 2,15}$--16$_{\rm 1,15}$) o1 line ($E_{\rm u}$ = 327 K, $S_{ij} \mu^{2}$ = 8 $D^{2}$) is partially blended with the SO$_{2}$ (8 -- 7) transition. \\

\end{table}

\end{document}